\documentclass[journal,10pt]{IEEEtran}
\usepackage{amsfonts}
\IEEEoverridecommandlockouts

\ifCLASSINFOpdf
\else
\fi
\usepackage{epsfig}
\usepackage{graphicx}
\usepackage{subfigure}
\usepackage{psfig}
\usepackage{epsf}
\usepackage[cmex10]{amsmath}
\usepackage{booktabs}
\usepackage{fancyhdr}
\usepackage{color}
\usepackage{booktabs}
\usepackage{stfloats}
\begin{document}
\title{Robust Resource Allocation for MISO Cognitive Radio Networks Under Two Practical Non-Linear Energy Harvesting  Models}
\author{\IEEEauthorblockN{Xiongjian Zhang, Yuhao Wang, Fuhui Zhou, \emph{Member, IEEE}, Naofal Al-Dhahir, \emph{Fellow, IEEE}, Xiaohua Deng}

\thanks{X. Zhang, Y. Wang, F. Zhou and X. Deng are with Nanchang University, P. R. China (e-mail:  zhangxiongjian@email.ncu.edu.cn, wangyuhao@ncu.edu.cn, zhoufuhui@ieee.org). F. Zhou is also with  the Postdoctoral Research Station of Environment Science and
Engineering, Nanchang University. N. Al-Dhahir is with University of Texas at Dallas, USA. (e-mail: aldhahir@utdallas.edu).  The corresponding author is Fuhui Zhou.

The research was supported in part by the Natural Science Foundation of China (61701214, 61261010 and 61661028), in part by the Graduate Student Innovation and Entrepreneurship Project of Jiangxi Province (CX2017196), in part by the Young Natural Science Foundation (20171BAB212002), in part by The Postdoctoral Science of Jiangxi Province (2017M610400, 2017KY04 and 2017RC17), in party by the key Project for Young Natural Science of Jiangxi Province (20152ACB21008), and in part by the Young Scientist of Jiangxi province (20142BCB2300). }
}
\maketitle
%Department of Electrical and Computer Engineering as a Research Fellow at
\begin{abstract}
To achieve a good tradeoff between the consumed power and the harvested power, a robust resource optimization problem is studied in a multiple-input single-output cognitive radio network with simultaneous wireless information and power transfer under imperfect channel state information. Unlike most of the existing works that assume an ideal linear energy harvesting (EH) model, we assume two practical non-linear EH models. In order to solve the resulting challenging non-convex problem, we propose an algorithm based on successive convex approximation and prove that a rank-one solution is obtained in each iteration of the proposed algorithm. Our simulation results quantify the effect of the sensitivity threshold of the EH circuit on the harvested power.
\end{abstract}
\vbox{} %空出一行命令
\IEEEpeerreviewmaketitle
\section{Introduction}
\subsection{Motivation and Related Work}
The application of  simultaneous wireless information and power transfer (SWIPT) to CRNs is a promising solution to simultaneously improve the spectrum efficiency (SE) and energy efficiency \cite{I. Krikidis}-\cite{C. Xu}. Up to now, the existing SWIPT literature can be categorized into two research directions based on the energy harvesting (EH) model, namely, the linear EH model \cite{F. Zhou}-\cite{C. Xu} and the non-linear EH model \cite{E. Boshkovska}-\cite{Y. Wang}. In \cite{F. Zhou}, we studied robust resource allocation in CRNs with SWIPT and proposed optimal beamforming schemes for minimizing the transmission power of the CBS. In \cite{D. W. K. Ng}, the authors proposed a multi-objective optimization problem for MISO CRNs with SWIPT. It was shown that there are several tradeoffs among the objectives considered in \cite{D. W. K. Ng}, such as the tradeoff between the harvested energy and the secrecy rate. To further improve SE and energy transfer efficiency, the authors studied the precoding design problem in multiple-input multiple-output CRNs with SWIPT \cite{C. Xu}.

Although the beamforming design problems have been well studied in CRNs with SWIPT under the ideal linear EH model \cite{Z. Chu}-\cite{C. Xu}, the proposed beamforming schemes are not suitable or suboptimal in practice due to the non-linear behavior of practical EH circuits \cite{E. Boshkovska}-\cite{Y. Wang}. In order to address this issue, the authors in \cite{E. Boshkovska} proposed a practical non-linear EH model and a beamforming scheme for maximizing the total harvested power of all EHRs. It was shown that the beamforming scheme designed under the non-linear EH model and achieves better performance than that achieved under the linear EH model. These investigations were extended to MIMO systems in \cite{E. Boshkovska2}-\cite{K. Xiong}. More recently, the authors in \cite{Y. Huang} and \cite{Y. Wang} have proposed optimal resource allocation schemes in CRNs with SWIPT under the non-linear EH model.

However, the beamforming design problems in \cite{E. Boshkovska}-\cite{Y. Wang} were formulated in a single objective framework. It over-emphasizes the importance of one metric and cannot achieve a good tradeoff among multiple metrics, especially when there are several conflicting metrics, such as energy efficiency and SE. Thus, in order to achieve a good tradeoff between the transmission power of the CBS and the harvested power of EHRs, unlike the work in \cite{E. Boshkovska}-\cite{Y. Wang}, an optimization problem with a weighted sum of multiple objectives is established for MISO CRNs with SWIPT under two non-linear EH models.
\subsection{Contributions and Organization}
The main contributions of our work are: Unlike the existing works in \cite{Z. Chu}-\cite{C. Xu}, a robust resource allocation scheme is proposed for CRNs with SWIPT under two practical non-linear EH models and imperfect CSI. The successive convex approximation (SCA) technique and ${\cal S}\text{-Procedure} $ are exploited to solve the formulated challenging non-convex problems. We prove that the solutions obtained using SCA satisfy the Karush-Kuhn-Tucker (KKT) conditions. In addition, our simulation results compare the performances achieved under the two non-linear EH models.

The remain of this letter is organized as follows. Section II presents the system model. The robust resource allocation problem is formulated in Section III. Section IV presents simulation results and Section V concludes this letter.

\section{System Model}
In this letter, a downlink MISO CR network with SWIPT is considered, where the primary network coexists with the secondary network under the spectrum sharing paradigm \cite{D. W. K. Ng}. The primary network consists of a primary base station (PBS) and $J$ primary users (PUs). In the secondary network, the CBS transmits information to a secondary user (SU) and simultaneously transfers energy to $K$ EHRs. Only one secondary user is assumed in order to reduce the multiple access interference caused to the primary user receivers. This assumption has been widely used \cite{F. Zhou}, \cite{D. W. K. Ng}.  The PBS is equipped with $N_p$ ($N_p>1$) antennas while the CBS is equipped with $N_t$ ($N_t>1$) antennas. The SU, PUs and EHRs are each equipped with a single antenna. The signals received at the SU, the $k$th $(k=1,2,...,K)$ EHR, and the interference signals affecting the $j$th $(j=1,2,...,J)$ PU from the CBS are denoted by $y$, $y_{e,k}$ and $y_{PR,j}$, respectively, and given by
\begin{subequations}
\begin{align}
&{y}={\textbf{h}^{\dagger}}{\textbf{w}_s}{x_s}+{\textbf{q} _{ps}^{\dagger}}{\sum\limits_{j=1}^J}{\textbf{w}_{p,j}}{x_p}+{n_s}\\
&{y_{e,k}}={\textbf{g}_{k}^{\dagger}{\textbf{w}_s}{x_s}}+{\textbf{q}_{pe,k}^{\dagger}}{\sum\limits_{j=1}^J}{\textbf{w}_{p,j}}{x_p}+{n_{e,k}}\\
&{y_{PR,j}}={\textbf{e}_{j}^{\dagger}}{\textbf{w}_s}{x_s}
\end{align}
\end{subequations}
where $\textbf{h} \in \textbf{C}^{N_t\times1}$, $\textbf{g}_{k} \in \textbf{C}^{N_t\times1}$ and $\textbf{e}_{j} \in \textbf{C}^{N_t\times1}$ denote the channel vectors between the CBS and the SU, the $k$th EHR and the $j$th PU, respectively. Boldface lower case letters represent vectors;
$\mathbf{x}^\dagger$ represents the conjugate transpose of a vector $\mathbf{x}$;
$\textbf{C}^{M \times N}$ represents a $M$-by-$N$ dimensional complex matrix set.
In addition, $\textbf{q} _{ps} \in \textbf{C}^{N_p\times1}$ and $\textbf{q} _{pe,k} \in \textbf{C}^{N_p\times1}$ are the channel vectors from the PBS to the SU and the $k$th EHR, respectively.
In (1), $\textbf{w}_s \in \textbf{C}^{N_t\times1}$ and $\textbf{w}_{p,j} \in \textbf{C}^{N_p\times1}$, respectively, denote the corresponding beamforming vectors of the CBS and the PBS while $n_s \sim \mathcal{C} \mathcal{N} (0,\sigma_{s}^2)$ and $n_{e,k} \sim \mathcal{C} \mathcal{N} (0,\sigma_{e,k}^2)$ denote the zero-mean additive white Gaussian noise (AWGN) at the SU and the $k$th EHR, respectively.
In (1), $x_s$ and $x_p$ are the data symbols intended for the SU and the PUs, respectively. Without loss of generality, it is assumed that $\mathbb{E}[xx^{\dagger}]=1$. Similar to \cite{F. Zhou}-\cite{C. Xu}, the PBS adopts a constant power transmission policy.

A practical spectrum sharing  scenario where there is no cooperation among the CBS, PUs and EHRs due to the limited resource is considered. In this case, $\textbf{g}_{k}$, $\textbf{e}_{j}$, $\textbf{q} _{ps}$ and $\textbf{q} _{pe,k}$ are imperfect and modeled as $\textbf{g}_{k} = {\overline{\textbf{g}}  _k} + \Delta {\textbf{g}_k}$, ${\textbf{e}_j} = {\overline{ \textbf{e}} _j} + \Delta {\textbf{e}_j}$, ${\textbf{q} _{ps}} = {\overline {\textbf{q}} _{ps}} + \Delta {\textbf{q}_{ps}}$ and ${\textbf{q} _{pe,k}} = {\overline {\textbf{q}} _{pe,k}} + \Delta {\textbf{q} _{pe,k}}$, respectively \cite{F. Zhou}-\cite{C. Xu}. ${\overline{\textbf{g}}  _k} $, ${\overline{ \textbf{e}} _j}$, ${\overline {\textbf{q}} _{ps}}$ and ${\overline {\textbf{q}} _{pe,k}}$ are the estimates of $\textbf{g}_{k}$, $\textbf{e}_{j}$, $\textbf{q} _{ps}$ and $\textbf{q} _{pe,k}$, and $\Delta {\textbf{g}_k}$, $\Delta {\textbf{e}_j}$, $\Delta {\textbf{q}_{ps}}$, and $\Delta {\textbf{q} _{pe,k}}$ are the corresponding  estimation errors, where ${\mathbf{\Psi} _{g,k}} \buildrel \Delta \over = \left\{ {\Delta {\mathbf{g}_k} \in {\mathbf{C}^{{N_t} \times 1}}:\Delta \mathbf{g}_k^\dag \Delta {\mathbf{g}_k} \le \xi _{g,k}^2} \right\}$,
${\mathbf{\Psi} _{e,j}} \buildrel \Delta \over = \left\{ {\Delta {\mathbf{e}_j} \in {\mathbf{C}^{{N_t} \times 1}}:\Delta \mathbf{e}_j^\dag \Delta {\mathbf{e}_j} \le \xi _{e,j}^2} \right\}$,
${\mathbf{\Psi} _{ps}} \buildrel \Delta \over = \left\{ {\Delta {\mathbf{q}_{ps}} \in {\mathbf{C}^{{N_p} \times 1}}:\Delta \mathbf{q}_{ps}^\dag \Delta {\mathbf{q}_{ps}} \le \xi _{ps}^2} \right\}$ and ${\mathbf{\Psi} _{pe,k}} \buildrel \Delta \over =$ $ \left\{ {\Delta {\mathbf{q}_{pe,k}} \in {\mathbf{C}^{{N_p} \times 1}}:\Delta \mathbf{q}_{pe,k}^\dag \Delta {\mathbf{q}_{pe,k}} \le \xi _{pe,k}^2} \right\}$ denote the uncertainty regions, respectively.
$\xi_{g,k}$, $\xi_{e,j}$, $\xi_{ps}$ and $\xi_{pe,k}$ are the corresponding radii of the uncertainty regions.

%$ {\Delta \mathbf{g}_k^\dag \Delta {\mathbf{g}_k} \le \xi _{g,k}^2}$, $  {\Delta \mathbf{e}_j^\dag \Delta {\mathbf{e}_j} \le \xi _{e,j}^2}$, ${\Delta \mathbf{q}_{ps}^\dag \Delta {\mathbf{q}_{ps}} \le \xi _{ps}^2}$ and ${\Delta \mathbf{q}_{pe,k}^\dag \Delta {\mathbf{q}_{pe,k}} \le \xi _{pe,k}^2}$.

The total radio frequency (RF) power received at the $k$th EHR denoted by $P_{ER,k}$, can be represented as
\begin{align}
&{P_{ER,k}}= {\text{Tr}( {\textbf{g}_k^{\dagger}}{\textbf{w}_s}{\textbf{w}_s^{\dagger}}{\textbf{g}_k}}) + { \sum\limits_{j=1}^J \text{Tr} ({\textbf{q}_{pe,k}^{\dagger}}{\textbf{w}_{p,j}}{\textbf{w}_{p,j}^{\dagger}} {\textbf{q}_{pe,k}})}.
\end{align}

In this letter, two practical non-linear EH models are considered.
According to \cite{E. Boshkovska} and \cite{E. Boshkovska2}, the harvested energy in the first model at the $k$th EHR denoted by $\Phi_{ER,k}$ is given by
\begin{subequations}
\begin{align}
&{\Phi_{ER,k}}=\frac {\Psi_{ER,k}-{M_k}{\Omega_k}} {1-\Omega_k}, {\Omega_k}=\frac {1} {1+\exp\left({a_k}{b_k}\right)}\\
&{\Psi_{ER,k}}=\frac {M_k} {1+\exp{\left(-a_k\left({P_{ER,k}-b_k}\right)\right)}}
\end{align}
\end{subequations}
where $M_k$ represents the maximum harvested power at the $k$th EHR when the EH circuit is satured. Parameters $a_k$ and $b_k$ are related to the specifications of the EH circuit.

Recently, another practical non-linear EH model has been proposed in \cite{S. Wang}, where the authors considered a sensitivity property in the non-linear EH model. More specifically, the harvested energy is set to zero when the input RF power is smaller than the EHRs' sensitivity threshold. The non-linear EH model proposed in \cite{S. Wang} is given by
\begin{align}
&{\Theta_{ER,k}}=[\frac {M_k} {\exp(-c P_{0} + n)} (\frac {1+\exp(-c P_{0} + n)} {1+\exp(-c P_{ER,k} + n)}-1)]^{+}
\end{align}
where $P_0$ is the sensitivity threshold. Based on this sensitivity property, when $P_{ER,k}$ is smaller than $P_0$, $\Theta_{ER,k}$ is 0. The parameters $c$ and $n$ control the steepness of this function.

\section{Robust Resource Optimization Formulation}
To achieve a good tradeoff between the transmission power and the total harvested energy, an optimization problem with a weighted sum of multiple objectives is formulated. Let $\textbf{W}={\textbf{w}_s}{\textbf{w}_s^{\dagger}}$, $\textbf{W}_{p}=\sum\limits_{j=1}^{J}{\textbf{w}_{p,j}}{\textbf{w}_{p,j}^{\dagger}}$ and $\textbf{H}=\textbf{h} \textbf{h}^{\dagger}$.
Since $\Omega_k$ does not affect the beamforming design, $\Psi_{ER,k}$ can represent the harvested power at the $k$th EHR \cite{E. Boshkovska}.
The robust resource optimization problem can be formulated as
\begin{subequations}
\begin{align}
&\text{P}_1:\max_{\textbf{W}} \alpha\sum_{k=1}^{K} \Psi_{ER,k} - (1-\alpha)\text{Tr}(\textbf{W})\\
&\text{s.t.} \text{   } \text{    }     {C1:} {\textbf{e}^\dag_j\textbf{W}\textbf{e}_j \leq P_{In,j}}, \forall \Delta \textbf{e}_j \in \mathbf{\Psi} _{e,j}, \forall j\\
&{C2:} {\frac {\text{Tr}\left(\textbf{HW}\right)} {\textbf{q}^\dag_{ps} \textbf{W}_{p}\textbf{q}_{ps}+\sigma_s^2} \geq {\Gamma_{req}}}, \forall \Delta \textbf{q}_{ps} \in \mathbf{\Psi} _{ps}\\
&{C3:} {\text{Tr} \left(\textbf{W}\right) \leq P_{max}}\\
&{C4:} {\textbf{W} \succeq 0}\\
&{C5:} {\text{Rank} \left(\textbf{W}\right) = 1}
\end{align}
\end{subequations}
where the constant $\alpha$ is a weight in the range $0\leq \alpha \leq 1$.
In (5a), $\alpha$ and $1-\alpha$ indicate the preference of the system designer for the harvested energy maximization and the transmit power minimization, respectively. In the special case, when $\alpha=1$ or $\alpha=0$, P$_1$ is equivalent to the harvested energy maximization or the transmit power minimization, respectively, and $P_{In,j}$ is the maximum tolerable interference power of the $j$th PU. The constant $\Gamma_{req}$ is the minimum signal-to-interference-noise (SINR) required at the SU. The constraint $C2$ guarantees that the SINR of the SU is not less than $\Gamma_{req}$ and $P_{max}$ denotes the maximum transmit power of the CBS.

%To solve the formulated MOOP, a weighting approach is followed.
%Thus, the MOOP can be transformed into a single objective optimization problem given as
%\begin{subequations}
%\begin{align}
%&\text{P}_2:  \max_{\textbf{W}} \alpha\sum_{k=1}^{K} \Psi_{ER,k} - (1-\alpha)\text{Tr}(\textbf{W}) \\
%&\text{s.t.} \text{   } \text{    } C1-C5
%\end{align}
%\end{subequations}
%where the constant $\alpha$ is a weight in the range $0\leq \alpha \leq 1$.
%In (6a), $\alpha$ and $1-\alpha$ indicate the preference of the system designer for the harvested energy maximization and the transmit power minimization, respectively. In the special case, when $\alpha=1$ or $\alpha=0$, P$_2$ is equivalent to the harvested energy maximization or the transmit power minimization, respectively.

It is observed that the objective function in P$_1$ is non-convex. To this end, we introduce auxiliary variables ($\tau_1, \tau_2, ... \tau_k$) to transform the non-convex objective function into an equivalent convex form.
Concretely, P$_1$ can be equivalently expressed as
\begin{subequations}
\begin{align}
&\text{P}_2:  \max_{\textbf{W},\tau_k} \alpha \sum\limits_{k=1}^{K} \tau_{k}-(1-\alpha) \text{Tr} (\textbf{W}) \\
&\text{s.t.} C6:\frac {M_k} {1+\exp{\left(-a_k\left({P_{ER,k}-b_k}\right)\right)}} \geq \tau_k,\\
&\ \ \ \ \ \forall \Delta \textbf{g}_k \in \mathbf{\Psi}_{g,k}, \forall \Delta \textbf{q}_{pe,k} \in \mathbf{\Psi}_{pe,k}, \forall k, \text{   }  C1-C5. \nonumber
\end{align}
\end{subequations}

Although the transformed objective function is linear, the optimization problem P$_2$ is still non-convex due to the rank-one constraint $C5$ and the non-convex constraint $C6$. To obtain a tractable solution, semi-definite relaxation (SDR) is exploited to tackle the constraint $C5$ and the SCA technique is applied to address $C6$. Thus, P$_2$ can be solved by iteratively solving P$_3$ given by
\begin{subequations}
\begin{align}
&\text{P}_3:  \max_{\textbf{W},\tau_{k}} \alpha \sum\limits_{k=1}^{K} \tau_{k}-(1-\alpha) \text{Tr} (\textbf{W}) \\
&\text{s.t.} \text{   } \text{    } C1-C4\\
&\overline{C6}: -{a_k} {\textbf{g}^\dag_k} {\textbf{W}} \textbf{g}_k \leq \ln{\left( {M_k}-\tau_{k} \right)}-\ln{\tau_{m,k}} -\\ \nonumber
&\text{   }\text{   }\text{   }\text{   }\text{   }\text{   }  \frac{\tau_{k}-\tau_{m,k}}{\tau_{m,k}}+{a_k} \textbf{q}^\dag_{pe,k}{\textbf{W}_p} \textbf{q}_{pe,k}- a_k b_k,\\  \nonumber
&\ \ \ \ \ \forall \Delta \textbf{g}_k \in \mathbf{\Psi}_{g,k}, \forall \Delta \textbf{q}_{pe,k} \in \mathbf{\Psi}_{pe,k}, \forall k.
\end{align}
\end{subequations}
where $\tau_{m,k}$ is a constant at the $m$th iteration of the proposed algorithm, which is shown in Table I and $m$ denotes the number of iterations. In order to address the channel uncertainty, the ${\cal S}\text{-Procedure}$ is applied \cite{F. Zhou}.
%It is seen that P$_4$ is convex and can be easily solved by using convex optimization tools such as CVX \cite{F. Zhou}.
Let $\mathbf{O}_{e,j}=[\textbf{I} \text{ } \overline {\textbf{e}}_{j}]$, ${\mathbf{\Upsilon}_{q} = [\textbf{I} \text{ } \overline {\textbf{q}}_{ps}]}$, $\mathbf{O}_{g,k}=[\textbf{I} \text{ } \overline {\textbf{g}}_{k}]$ and ${\mathbf{\Upsilon}_{q,k} = [\textbf{I} \text{ } \overline {\textbf{q}}_{pe,k}]}$. Thus, the constraints (5b), (5c) and (7c) can be equivalently expressed as
\begin{align}
&\mathbf{\Gamma}(\mu_j,\textbf{W})= \begin{bmatrix} \mu_j\textbf{I} &\textbf{ 0}\\ \textbf{ 0} & P_{In,j}-\mu_j\xi^2_{e,j}\end{bmatrix}-\mathbf{O}^\dag_{e,j} \textbf{W} \mathbf{O}_{e,j} \succeq 0
\end{align}
\begin{align}
&\mathbf{\Gamma}(\varpi_q,\textbf{W})=\begin{bmatrix} \varpi_q \textbf{I} & \textbf{0}\\ \textbf{ 0} & \frac{\text{Tr}(\textbf{HW})}{\Gamma_{req}}-\sigma^2_s-\varpi_q \xi^2_{ps} \end{bmatrix}-\mathbf{\Upsilon}^\dag_q \textbf{W}_p \mathbf{\Upsilon}_q \succeq 0
\end{align}
\begin{subequations}
\begin{align}
&\mathbf{\Gamma}_{k}(\varpi_{g,k},\delta_{q,k},\tau_k,\textbf{W})= \\
&\begin{bmatrix} \varpi_{g,k} \textbf{I} & \textbf{ 0} \\ \textbf{ 0} & \delta_{q,k}-\Lambda_k-\varpi_{g,k} \xi^2_{g,k} \end{bmatrix}+\mathbf{O}^\dag_{g,k} \textbf{W} \mathbf{O}_{g,k} \succeq 0 \nonumber \\
&\mathbf{\Gamma}_{k}(\beta_{q,k},\delta_{q,k})=\\
&\begin{bmatrix} \beta_{q,k} \textbf{I} & \textbf{ 0} \\ \textbf{ 0} & -\delta_{q,k}-\beta_{q,k} \xi^2_{pe,k} \end{bmatrix}+\mathbf{\Upsilon}^\dag_{q,k} \textbf{W}_p \mathbf{\Upsilon}_{q,k} \succeq 0 \nonumber
\end{align}
\end{subequations}
where $\Lambda_k= \text{ln}\left(\tau_{m,k}/a_k\right) + (\tau_k-\tau_{m,k})/a_k\tau_{m,k}+b_k-\text{ln}(M_K-\tau_k)/a_k$, $\mu_j \geq 0$, $\varpi_q \geq 0$, $\varpi_{g,k} \geq 0$, $\beta_{q,k} \geq 0$ and $\delta_{q,k}$ are slack variables. It is seen from $C6$ that $0\leq\tau_k\leq M_k$. Thus, $\tau_k$ can be obtained by using one dimensional search method. In this case, P$_3$  can be solved by solving P$_4$ given as
\begin{subequations}
\begin{align}
&\text{P}_4:  \max_{\textbf{W},\Xi} \alpha \sum\limits_{k=1}^{K} \tau_{k}-(1-\alpha) \text{Tr} (\textbf{W}) \\
&\text{s.t.}\ \ \text{(5d)}, \text{(5e)}, \text{(8)}, \text{(9)}, \text{(10)}
\end{align}
\end{subequations}
where $\Xi$ is the set of all slack variables. It is seen that P$_4$ is convex and can be easily solved by using convex optimization tools such as CVX \cite{F. Zhou}.

\emph{\textbf{Theorem 1:}} For problem P$_4$, it is assumed that $\Gamma_{req} \geq 0$. The optimal \textbf{W} matrix is unique and its rank is one.

\emph{\textbf{Proof:}} See Appendix A.

\emph{\textbf{Theorem 2:}} When the non-linear EH model given by eq. (4) is adopted, the robust resource optimization problem can be also solved by using Algorithm 1 and the optimal \textbf{W} matrix is unique and rank-one.

%\emph{\textbf{Theorem 1:}} For problem P$_4$, it is assumed that $\Gamma_{req} \geq 0$. The optimal \textbf{W} matrix is unique and its rank is one.

%\emph{\textbf{Proof:}} See Appendix A.

%Based on Theorem 1, an iterative algorithm denoted by Algorithm 1 is proposed for solving P$_2$. The details of Algorithm 1 are given in Table I.

%\emph{\textbf{Theorem 2:}} When the non-linear EH model given by Eq. (4) is adopted, the MOOP can be also solved by using Algorithm 1 and the optimal \textbf{W} matrix is unique and rank-one.

%\emph{\textbf{Proof:}} The proof is similar to the proof of Theorem 1 and it is omitted due to space limitation.

\begin{table}[htbp]
\begin{center}
\caption{The SCA-based algorithm}
\begin{tabular}{lcl}
\\\toprule
$\textbf{Algorithm 1}$: The iterative algorithm for $\text{P}_{{2}}$\\ \midrule
\  1: \textbf{Setting:}\\
 \ \ \ \ \ \ \ $P_{In,j}, \Gamma_{req}, P_{max}, M_k, \xi_{e,j}, \xi_{ps}, \xi_{g,k}, \xi_{pe,k}$.\\
 \ 2: \textbf{Inputting:}\\
 \ \ \ \ \ \ \ $\textbf{w}_{p,j}$, $\alpha$, the CSI $\textbf{h}$, the estimated CSI ${\overline{\textbf{g}}  _k} $, ${\overline{ \textbf{e}} _j}$, ${\overline {\textbf{q}} _{ps}}$ and ${\overline {\textbf{q}} _{pe,k}}$.\\
 \ 3: \textbf{Initialization:}\\
 \ \ \ \ \ \ \ the iteration index $m=1$, the maximum accuracy $\epsilon$,\\
 \ \ \ \ \ \ \ $\tau_k$'s initialization $\tau_0$, one-dimensional search step $s$.\\
 \ 4: \textbf{Optimization:}\\
 \ \ \ \ \ \ \ \textbf{for} $\tau_k=\tau_0:s:M_k$\\
 \ \ \ \ \ \ \ \ \ \ \textbf{repeat} \\
 \ \ \ \ \ \ \ \ \ \ \ \ Solve the problem P$_4$ by the software CVX, obtain the\\
 \ \ \ \ \ \ \ \ \ \ \ \ objective value $\Sigma_{m}$ and the intermediate beamformer $\textbf{w}$,\\
 \ \ \ \ \ \ \ \ \ \ \ \ where $\Sigma_m=\alpha\sum\limits_{k=1}^{K} \tau_{k}-(1-\alpha) \text{Tr} (\textbf{W})$.\\
 \ \ \ \ \ \ \ \ \ \ \ \ \textbf{if} $\Sigma_{m}-\Sigma_{m-1} \leq \epsilon$ is satisfied, then\\
 \ \ \ \ \ \ \ \ \ \ \ \ \ \ \ \ \ \ \textbf{return} the objective value $\Sigma=\Sigma_{m}$ and $\textbf{w}$.\\
 \ \ \ \ \ \ \ \ \ \ \ \ \textbf{else}\\
 \ \ \ \ \ \ \ \ \ \ \ \ \ \ \ \ \ \ Update $\tau_{m,k}$ to $\tau_{m+1,k}$ according to $\tau_k = \Psi_{ER,k}$ \\
 \ \ \ \ \ \ \ \ \ \ \ \ \ \ \ \ \ \ and $m=m+1$.\\
 \ \ \ \ \ \ \ \ \ \ \ \ \textbf{end if}\\
 \ \ \ \ \ \ \ \ \ \ \textbf{until} $\Sigma_{m}-\Sigma_{m-1} \leq \epsilon$ is satisfied.\\
 \ \ \ \ \ \ \ \textbf{end}\\
 \ \ \ \ \ \ \ \textbf{return} the max $\Sigma$ and the corresponding $\textbf{w}$, obtain $\textbf{w}^{*}=\textbf{w}$.
\\\bottomrule
\end{tabular}
\end{center}
\end{table}

\section{Simulation Results}
The simulation parameters are based on the works in \cite{E. Boshkovska} and \cite{S. Wang}, given as: $N_t=3$, $N_p=3$, $K=2$, $J=2$, $P_{In,j}=10$ dBm, $P_{max}=20$ dBm, $\sigma_{s}^2=\sigma_{e,k}^2=-120$ dBm, $a_k=150$, $b_k=0.014$, $c=150$, $n=2.1$, $P_0=6.4$ mW, $M_k=24$ mW, $ {\mathbf{h}} \sim {\cal C}{\cal N}\left( {0,{\mathbf{I}}} \right)$, ${\overline{\textbf{g}}  _k} \sim {\cal C}{\cal N}\left( {0,{0.5\mathbf{I}}} \right)$, ${\overline{ \textbf{e}} _j}\sim {\cal C}{\cal N}\left( {0,{\mathbf{I}}} \right)$, ${\overline {\textbf{q}} _{ps}}\sim {\cal C}{\cal N}\left( {0,{\mathbf{I}}} \right)$, ${\overline {\textbf{q}} _{pe,k}}\sim {\cal C}{\cal N}\left( {0,0.1{\mathbf{I}}} \right)$, $\xi_{g,k}=0.002$, $\xi_{e,j}=0.005$, $\xi_{ps}=0.005$ and $\xi_{pe,k}=0.001$. The number of channel realizations is $10^4$.

\begin{figure*}[!t]
\centering
\subfigure[The relationship among the harvested power, the transmit power and the minimum SINR requirement.] {\includegraphics[height=2in,width=2.5in,angle=0]{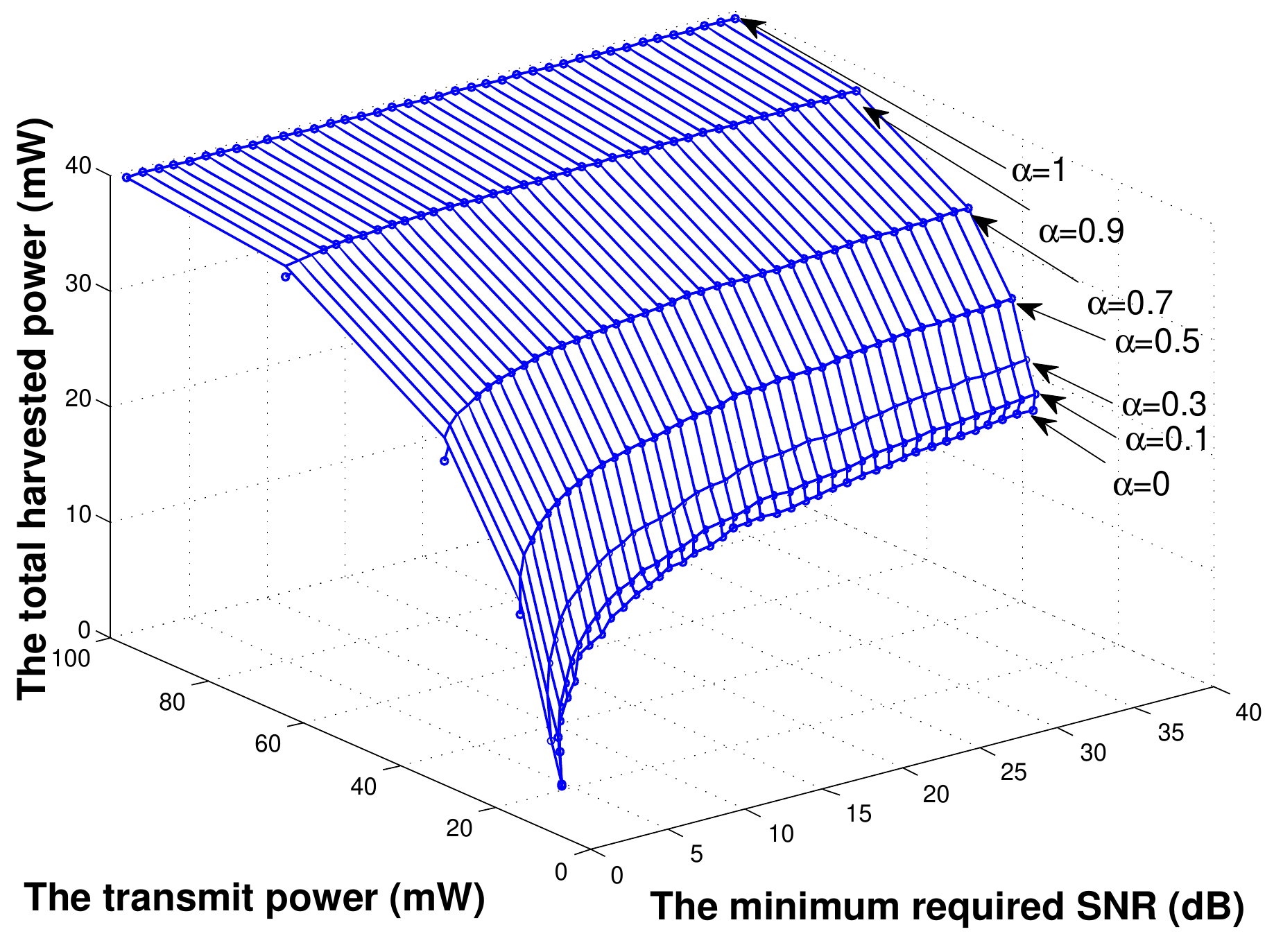}}
\subfigure[The total harvested power versus the minimum required SINR.] {\includegraphics[height=2in,width=2.2in,angle=0]{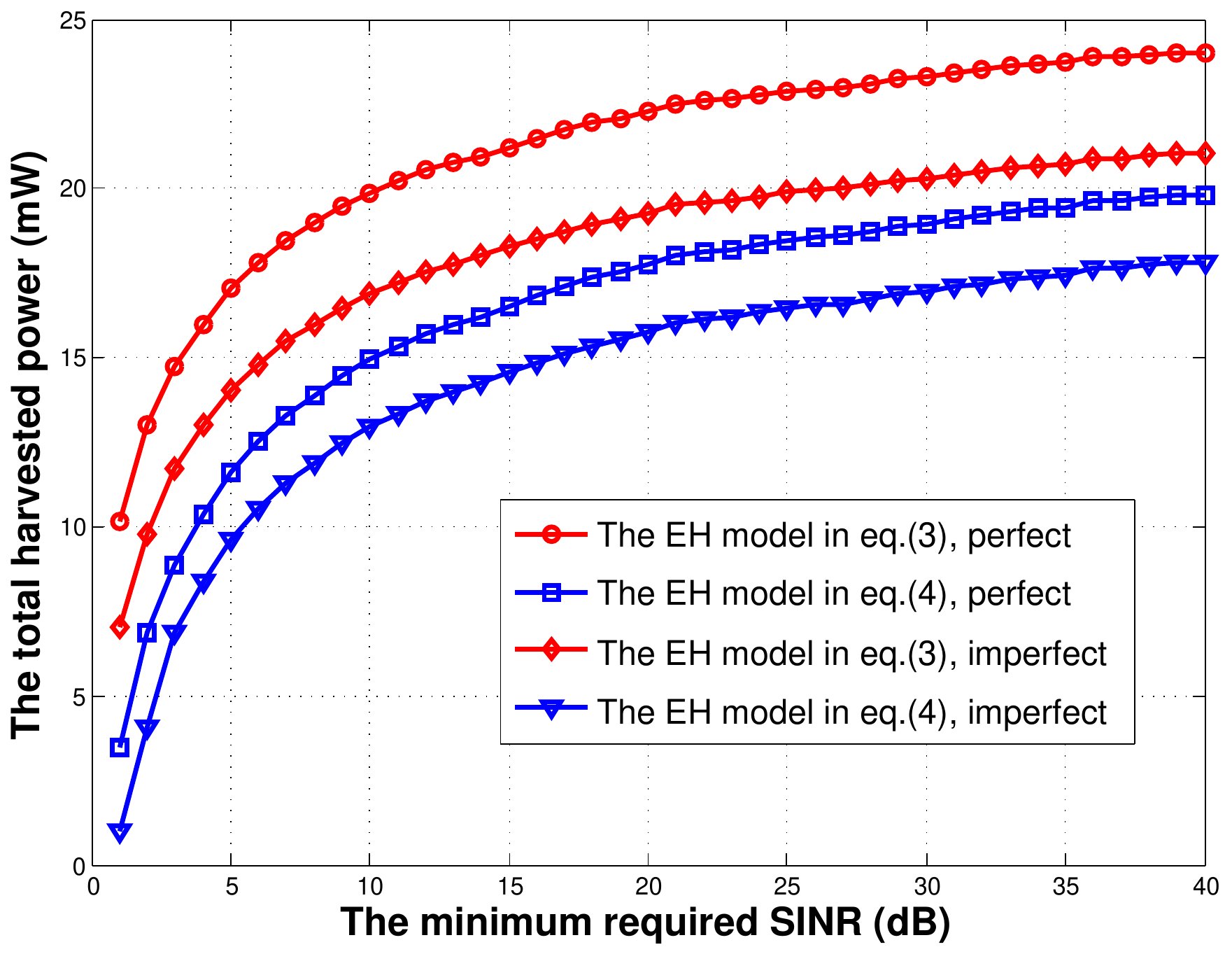}}
\subfigure[Values of the objective function versus the number of iterations with different values of $\alpha$.] {\includegraphics[height=2in,width=2.2in,angle=0]{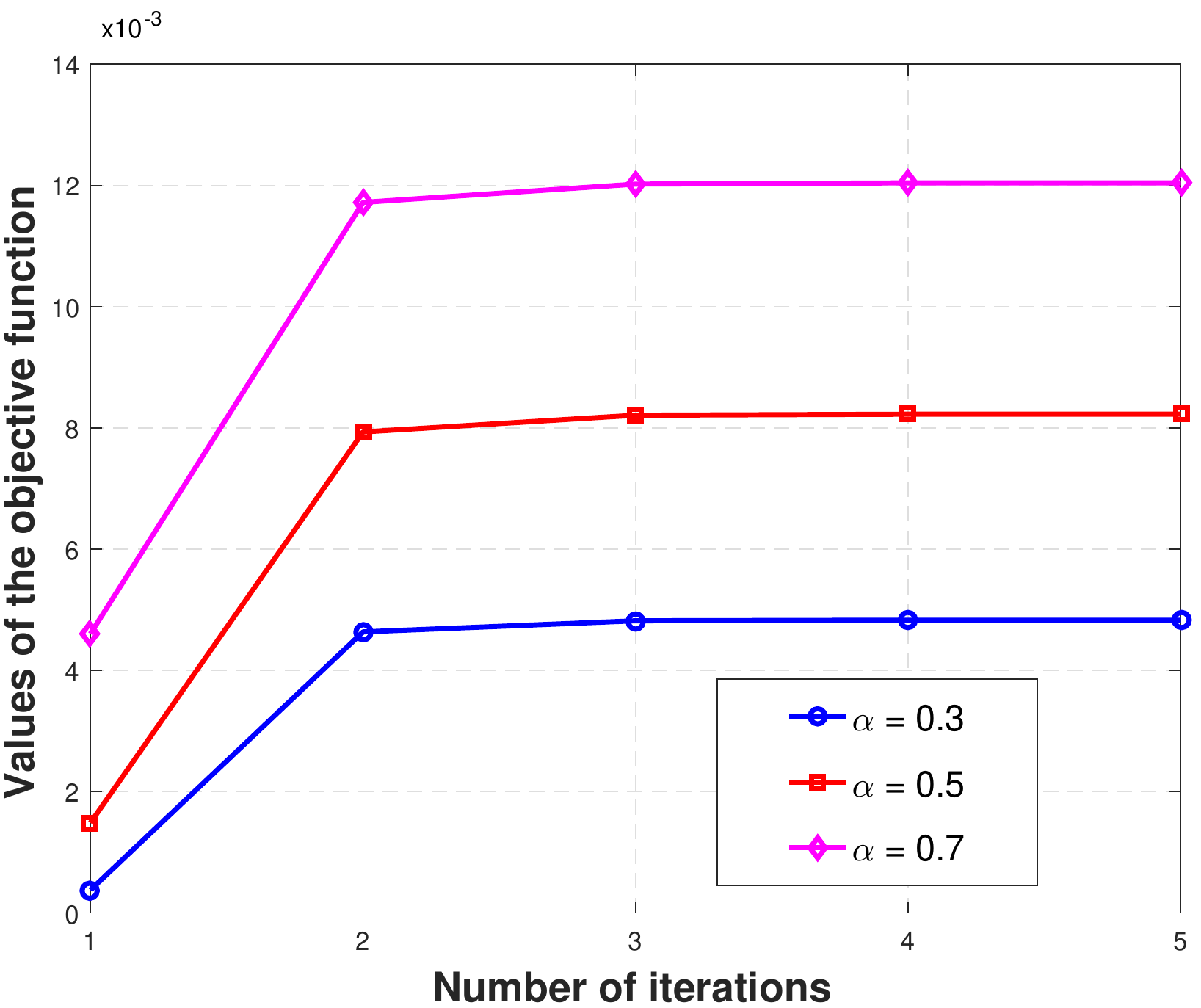}}
\label{fig2}
\end{figure*}
Fig. 1(a) depicts the tradeoff between the total harvested power and the transmit power achieved by using the proposed algorithm. The results are obtained with different weights $\alpha (\alpha=0, 0.1, 0.3, 0.5, 0.7, 0.9, 1)$.
It is seen that the total harvested power increases with the minimum required SINR for a fixed weight $\alpha$. When SINR is increased above a certain level (such as 35 dB), the harvested power does not change and becomes a constant. The reason is that the harvested power directly depends on the transmit power and the transmit power increases when the SINR is increased. However, the transmit power cannot be higher than the maximum transmit power.
On the other hand, for a fixed $\Gamma_{req}$, the total harvested power increases with the weight $\alpha$. The reason is that the effect of the harvested power increases with the weight $\alpha$.

Fig. 1(b) shows the total harvested power versus the minimum SINR required at the SU under different CSI conditions. It is seen from Fig. 1(b) that the performance achieved under the non-linear EH model given by eq. (3) is better than that obtained by using the model given by eq. (4) since the former model does not consider the sensitivity property of the EH circuit unlike the latter model. It is also seen that the imperfect CSI significantly influences the total harvested power since a higher transmit power is required to overcome the CSI uncertainty. Fig. 1(c) verifies the efficiency of our proposed algorithm 1 since it is seen that it only requires less than 4 iterations to converge.

\section{Conclusion}
A robust resource optimization problem was formulated in MISO CRNs with SWIPT where two practical non-linear EH models and imperfect CSI were investigated. An algorithm based on SCA  and ${\cal S}\text{-Procedure}$ was proposed to solve the challenging formulated problem. We proved that a rank-one solution is obtained at each iteration of the proposed algorithm. Our simulation results quantify the effect of the sensitivity threshold on the achievable performance. Comparing the performance under the two practical non-linear EH models provides valuable insights for the design of CRNs with SWIPT.

\begin{center}
\text{APPENDIX A}
\end{center}

The proof of Theorem 1 is based on the KKT optimal conditions of P$_4$. Let $\textbf{A}_j \in \mathbb{H}^N_+$, $\textbf{B}\in \mathbb{H}^N_+$, $\eta$, $\textbf{C}_k\in \mathbb{H}^N_+$, $\textbf{D}_k\in \mathbb{H}^N_+$ and $\textbf{Z}\in \mathbb{H}^N_+$ denote the dual variables with respect to (8), (9), (5d), (10) and (5e), respectively. $\mathbb{H}^N_+$ denotes a \emph{N}-by-\emph{N} dimensional Hermitian positive semidefinite matrix set.
The partial KKT conditions related to the proof are given as
\begin{subequations}
\begin{align}
&\tag{12a} (\alpha - 1 - \eta )\textbf{I} +\mathbf{\Omega}- \frac{b_{nn}}{\Gamma_{req}}\textbf{H}- \textbf{Z} = 0 \nonumber\\
&\tag{12b} \textbf{Z} \textbf{W} = 0\\
&\tag{12c} \textbf{A}_j, \textbf{B}, \textbf{C}_k, \textbf{Z}, \textbf{W} \succeq 0, \alpha, \eta, b_{nn}, \Gamma_{req} \geq 0
\end{align}
\end{subequations}
where $\mathbf{\Omega}=\sum\limits_{j=1}^{J} \textbf{O}_{e,j} \textbf{A}_j \textbf{O}_{e,j}^\dag-\sum\limits_{k=1}^{K}\textbf{O}_{g,k}\textbf{C}_k \textbf{O}_{g,k}^\dag$, $b_{nn}$ is the element of \textbf{B} in the $n$th row of  the $n$th column.

Right-multiplying (12a) by $\textbf{W}$ and combining (12b), we get $\left(\left( \alpha-1- \eta \right) \textbf{I} +\mathbf{\Omega} \right) \textbf{W} = \frac{b_{nn}}{\Gamma_{req}}\textbf{H} \textbf{W}$, which implies that
\begin{align}
&\notag \text{Rank}[\left(\left( \alpha-1- \eta \right) \textbf{I} +\mathbf{\Omega} \right) \textbf{W}]\\
&\tag{13} =\text{Rank}(\frac{b_{nn}}{\Gamma_{req}}\textbf{H} \textbf{W}) \leq \text{Rank}(\textbf{H}) = 1.
\end{align}
Moreover, according to (12a) and (12c), one has
\begin{align}
&\tag{14} \left( \alpha-1- \eta \right) \textbf{I} +\mathbf{\Omega}=\textbf{Z}+ \frac{b_{nn}}{\Gamma_{req}}\textbf{H}.
\end{align}
Since $\textbf{Z}+ \frac{b_{nn}}{\Gamma_{req}}\textbf{H} \succ0$, one has $\left( \alpha-1- \eta \right) \textbf{I} +\mathbf{\Omega} \succ 0$. Thus,
\begin{align}
&\tag{15} \text{Rank}(\textbf{W}) = \text{Rank}[\left(\left( \alpha-1- \eta \right) \textbf{I} +\mathbf{\Omega} \right) \textbf{W}] \leq 1.
\end{align}
Based on (15), the rank of $\textbf{W}$ is one and the proof is complete.

\end{document}